\documentclass[notitlepage,showpacs,aps,pra,twocolumn]{revtex4-1}

\usepackage{amsmath,graphicx}

\renewcommand{\vector}[1]{\underline{\mathbf{#1}}} 
\newcommand{\eqnref}[1]{Eqn.~(\ref{#1})}
\newcommand{\figref}[1]{Fig.~\ref{#1}}

\begin{document}

\title{Retardation effects in induced atomic dipole-dipole interactions}

\author{S.D.~Graham and J.M.~McGuirk}
\affiliation{Department of Physics\\
Simon Fraser University, Burnaby, British Columbia V5A 1S6, Canada}
\date{\today}

\begin{abstract}
  We present mean-field calculations of azimuthally averaged retarded dipole-dipole interactions in a Bose-Einstein condensate induced by a laser, at both long and short wavelengths. Our calculations demonstrate that dipole-dipole interactions become significantly stronger at shorter wavelengths, by as much as $30$-fold, due to retardation effects. This enhancement, along with inclusion of the dynamic polarizability, indicate a method of inducing long-range interatomic interactions in neutral atom condensates at significantly lower intensities than previously realized.
\end{abstract}


\maketitle

\section{Introduction}
Bose-Einstein condensates (BECs) offer a theoretical and experimental platform for studying the physics of many-body systems. BECs can be used as easily manipulable testbeds for simulating many-body physics phenomena from condensed matter physics, quantum gases, and astrophysics. Of particular interest are systems where microscopic interatomic interactions give rise to macroscopic effects.

The most readily accessible atom-atom interactions in a BEC typically are the hard-sphere (s-wave) contact interactions, which, for instance, modify condensate ground-state shapes, perturb oscillation modes, and affect tunnelling rates in optical lattices \cite{Bloch2008}. This interaction can sometimes be tuned with magnetic fields via Feshbach resonances, where the interaction can change from repulsive to attractive \cite{Feshbach}. Such tunability is an important feature in generating models of many-body systems.

S-wave interactions, however, are isotropic and inherently local, limiting the range of accessible physical models. Dipole-dipole interactions, on the other hand, offer a different set of physical systems that cannot be studied with only the local s-wave interaction. These interactions, with their non-local components, have been demonstrated in atomic systems with permanent \cite{DDIExpansion} and induced dipoles \cite{DDRydberg}. In addition to the non-locality, dipole-dipole interactions are also anisotropic, with regions of attractive and repulsive interactions, which has been observed in the anisotropic expansion of dipolar BECs \cite{DDIExpansion} and scattering between Rydberg atoms \cite{RydbergAngular}.

There are several routes to realizing dipole-dipole interactions in ultracold gases. One method is to use atomic BECs with permanent dipole moments to explore the interplay between s-wave and dipolar interactions in many-body systems. Some groups have been successful in creating a dipolar BEC using atoms with permanent magnetic dipoles, such as Chromium \cite{ChrBEC}, Erbium \cite{ErbBEC}, and Dysprosium \cite{DysBEC}. These dipolar BECs have been used to explore new physics and macroscopic behaviour such as droplet states \cite{DropletFormation,DropletTheory} and roton dispersion \cite{RotonRadAndAng,RotonManifest,RotonLaser}. A second avenue to study dipole-dipole interactions comes from degenerate molecular gases, using molecules that have permanent electric dipole moments. Although molecular BECs have been created \cite{HMBEC,HMBEC2,MBEC,MBEC2}, the challenge and complexity of generating dense samples of ultra-cold heteronuclear molecules encourages an alternate approach.

Here, we consider a third approach: illuminating a neutral atomic BEC with an off-resonant laser to create induced dipoles, which then interact with each other. This approach leverages the robustness of atomic BEC creation and adds tunability, as the strength of these induced interactions may be tuned by adjusting the wavelength and intensity of the laser. Previous work suggested, though, that these interactions typically require unfeasibly enormous laser power to achieve comparable interaction strength to s-wave interactions \cite{SelfBindingBEC}. However, retardation effects from oscillating dipoles, which were not previously considered, may bring induced dipoles into the realm of reasonable power requirements. The strengthening of dipolar interactions due to retardation effects is possible in induced dipolar BECs, because dipole oscillations on the order of 100~THz or faster are required for propagating dipolar fields to be retarded significantly on the scale of BEC dimensions. Because retardation effects are generated by the oscillating fields that are inherent in creating the induced dipoles, retardation effects also could be observed by similarly driving oscillations in permanent dipoles at optical frequencies.

In previous work, a particularly compelling application of induced dipolar BECs was suggested. By using many different laser beams, a long-range $1/r$ potential could be induced in an atomic BEC. This gravitational-like interaction could be strong enough to self-bind a BEC and form a model for gravitationally-bound many-body systems, such as neutron stars \cite{OneOverRGravity}. This application is enticing, yet extremely complicated due to the multi-laser layout.

As an intermediate step to generating a self-bound BEC, a one-dimensional (1D) compression experiment would demonstrate the strength of induced dipole-dipole interactions. For 1D compression, a single laser beam is used to generate axial compression of a BEC via induced dipole-dipole interactions. Previous work used the variational principle to perform calculations of this system in the long-wavelength limit, where retardation effects are negligible \cite{OneD}. These results indicate that laser intensities of at least $10^8$~W/cm$^2$ are required to observe axial compression. However, at such large intensities the lifetime of a BEC would be reduced to $\sim 1$~ms, far too short to reach a stable ground state or observe dynamics. The enhancing effects of retarded interactions are needed to overcome these limitations.

In this paper, we present calculations that show retardation effects can be large for induced dipole-dipole interactions in BECs. These retardation effects are only present for interactions induced by short-wavelength lasers, where variational principle approaches break down and more complicated calculations are required. Retardation effects lead to an increase in interaction strength, lowering the required laser intensity and lengthening the lifetime of atoms in an induced dipolar BEC to more experimentally favorable values. Here, we first present the theory for induced dipole-dipole interactions, describe the numerical modeling techniques employed, present simulation results, and finally discuss the ramifications for the feasibility of observing induced dipole-dipole interactions.

\section{Theory: Azimuthally averaged laser-induced dipole-dipole interactions}
\begin{figure}
  \includegraphics[width=\linewidth,clip]{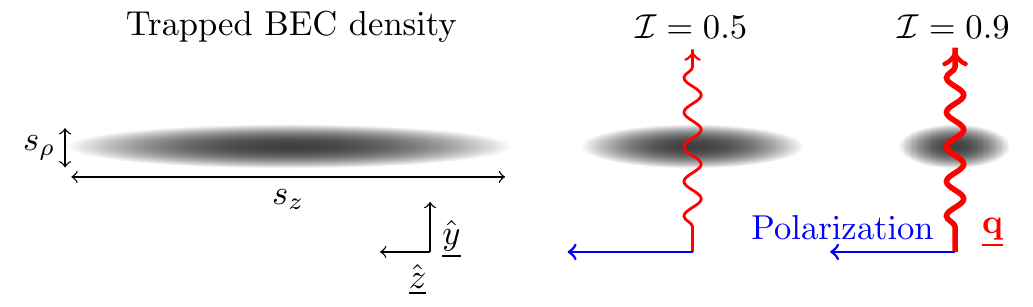}
  \caption{A linearly polarized laser beam illuminates a pencil-shaped condensate from the radial direction to induce dipole-dipole interactions. The condensate compresses axially as the relative laser intensity, $\mathcal{I}$, increases.}
  \label{fig:experiment}
\end{figure}

Following the approach of Ref.~\cite{OneD}, a trapped pencil-shaped BEC is illuminated with a uniform plane-wave laser polarized in the axial direction, shown in \figref{fig:experiment}. The laser induces electric dipoles in the atoms of the BEC, aligning the dipoles along the polarization axis of the laser. The choice of axial polarization suppresses superradiant Rayleigh scattering \cite{Superradiant} or collective atomic recoil lasing (CARL) \cite{CARL}, which are forbidden in the direction of polarization. The interaction potential between two atoms, separated by $\vector{r}$, with dipoles induced by a laser with wavevector $\vector{q}$ polarized in the $z$-direction is \cite{Thirunamachandran}
\begin{equation}
  \begin{split}
  U_{\text{DD}}(\vector{r}) = \frac{d^2}{r^3} \bigg[ \left( 1 - 3\cos^2\theta \right) \left( \cos(qr) + qr\sin(qr) \right) \\-  \left(\sin^2\theta\right) q^2r^2\cos(qr) \bigg] \cos(qy),
  \end{split}\label{eq:Udd}
\end{equation}
where $r$ is the interatomic distance, and $\theta$ is the angle between the interatomic axis and the polarization axis ($z$-axis), so that $\cos\theta = z/r$. The parameter $d^2 = \frac{I \alpha^2(q)}{4 \pi c \epsilon_0^2}$ is the induced dipole-dipole interaction strength; here, $I$ is the laser intensity, and $\alpha(q)$ is the dynamic atomic polarizability \cite{PolarizabilityBoninBook}. The tunable parameters for induced dipole-dipole interactions are the laser intensity and also frequency, since the polarizability is highly frequency-dependent.

The dipole-dipole interaction is three-dimensional in $(r, \theta, \phi)$, where $y = r \sin\theta \sin\phi$. We reduce the dimensionality by azimuthally averaging over $\phi$ to give a two-dimensional (2D) interaction,
\begin{equation}
  \begin{split}
  \bar{U}_{\text{DD}}(\vector{r}) = \frac{1}{2\pi} \int d\phi\, U_{\text{DD}} (\vector{r}) = \frac{d^2}{r^3} \bigg[  \left(1-3\cos^2\theta\right) \\
  \left( \cos(qr) + qr\sin(qr) \right) - \left(\sin^2\theta\right) q^2r^2\cos(qr) \bigg] J_0(q\rho).
  \end{split}\label{eq:Uddaa}
\end{equation}
Here $r^2 = z^2 + \rho^2$, and the trailing cosine in \eqnref{eq:Udd} has been converted to a Bessel function of the first kind, $J_0$. In the long-wavelength limit, this approximation reproduces analytic results from variational principle calculations \cite{OneD}. In the short-wavelength limit, the local condensate density is approximately isotropic and homogeneous in space on the scale of the laser wavelength, and this simplification is reasonable. A density-weighted azimuthal average would only marginally improve the accuracy of the calculations at the expense of significant computational resources.

In \eqnref{eq:Uddaa}, the retarded terms are those that are scaled by factors of $qr$ and $q^2r^2$, which are small in the long-wavelength regime ($qr \ll 1$). In this limit the instantaneous dipole-dipole interaction $\frac{d^2}{r^3} \left( 1-3\cos^2\theta \right)$ is reclaimed. However, in the short-wavelength regime ($qr > 1$) these retardation factors are large, amplifying the dipole-dipole interactions. Additionally, the dipole-dipole interaction strength is highly dependent on the atomic polarizability, which drastically increases near atomic resonances, by as much as $10^4$ times. These two effects -- atomic polarizability and retardation amplification -- are what make the short-wavelength regime desirable for demonstrating and studying dipole-dipole interactions.

The azimuthally averaged interaction will, if strong enough, alter the ground state of the BEC. To calculate this ground state, we begin with a mean-field approach. The mean-field Gross-Pitaevskii (GP) equation describing a BEC at zero temperature, with order parameter $\psi$ and non-local dipole-dipole interactions, is \cite{BECReview}
\begin{equation}
  \begin{split}
  \imath \hbar \frac{\partial \psi}{\partial t} = H \psi = \bigg[ - \frac{\hbar^2 \nabla^2}{2m} + V_{\text{ext}}(\vector{r}) + g n(\vector{r}) \\
  + \int d\vector{r}^\prime n(\vector{r}^\prime) \bar{U}_{\text{DD}}(\vector{r}^\prime - \vector{r}) \bigg] \psi,
  \end{split}\label{eq:GPE}
\end{equation}
where $|\psi(\vector{r},t)|^2 = n(\vector{r})$ is the BEC density, $V_{\text{ext}} = \frac{m}{2}\left( \omega_\rho^2 \rho^2 + \omega_z^2 z^2 \right)$ is the cylindrically symmetric trapping potential with trapping frequencies $\omega_\rho$ and $\omega_z$, and $g = 4 \pi \hbar^2 a/m$ is the s-wave interaction strength with scattering length $a$. The validity of the GP mean-field approximation is dependent on weak interactions \cite{OneOverRGravity}. This condition requires weak s-wave interactions, $n a^3 \ll 1$, as well as weak dipole-dipole interactions, $n a^3_{\text{dd}} \gg 1$, where $a_{\text{dd}} \simeq \hbar/md^2$. Both conditions are easily satisfied for small scattering lengths, $a$, and small dipole-dipole interaction strengths, $d^2$. However, following laser-induced collapse, these approximations can break down, as discussed later.

Evaluating the first three terms in the GP equation is straightforward; however, the dipole-dipole interaction is a computationally expensive convolution of density with $\bar{U}_{\text{DD}}$ that must be computed in the frequency domain. The convolution theorem gives
\begin{equation}
   \begin{split}
   \int d\vector{r}' n(\vector{r}') \bar{U}_{\text{DD}}(\vector{r}' - \vector{r}) = (2 \pi)^{3/2} \int d^3k\  e^{\imath \vector{k} \cdot \vector{r}}\,\hat{n}(\vector{k}) \hat{\bar{U}}_{\text{DD}}(\vector{k}) \\
   = (2 \pi)^{3} \mathcal{F}^{-1}_{\text{3D}}\{ \hat{n}(\vector{k}) \hat{\bar{U}}_{\text{DD}}(\vector{k})\},
   \end{split}
\end{equation}
where $\vector{k}$ is the cylindrically symmetric frequency coordinate, with components $k_\rho$ and $k_z$ in the radial and axial directions respectively. Additionally, $\hat{n}(\vector{k})$ and $\hat{\bar{U}}_{\text{DD}}(\vector{k})$ are the three-dimensional Fourier transforms ($\mathcal{F}_{\text{3D}}$) of the BEC density and the dipole-dipole interaction respectively.

\begin{figure}
  \includegraphics[width=\linewidth,clip]{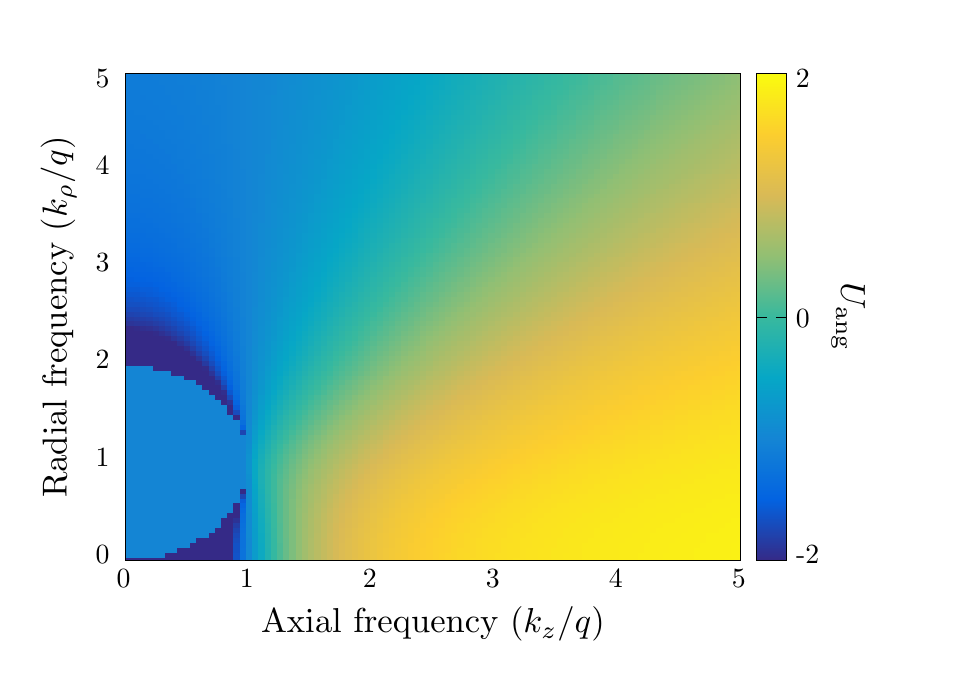}
  \caption{The dimensionless angular component of the spatial Fourier transform of azimuthally averaged induced dipole-dipole interactions. At small frequencies, a static attractive interaction dominates the long-range interaction. In the large frequency limit, the short-range instantaneous dipole interaction is dominant.}
  \label{fig:UDDFT}
\end{figure}

The Fourier transform of the azimuthally averaged dipole-dipole interaction, $\hat{\bar{U}}_{\text{DD}}(\vector{k}) = (2 \pi)^{-3/2}\int d^3r\  e^{-\imath \vector{k} \cdot \vector{r}}\, \bar{U}_{\text{DD}}(\vector{r})$, can be calculated using a similar technique as in Ref.~\cite{OneD}, giving
\begin{equation}
  \hat{\bar{U}}_{\text{DD}} (\vector{k}) = \frac{1}{(2\pi)^{3/2}}\frac{4\pi d^2}{3} U_{\text{ang}}(\vector{k}),
  \label{eq:UddaaFT}
\end{equation}
where the dimensionless angular component of the interaction is contained in
\begin{equation}
U_{\text{ang}}(\vector{k}) = -1 + 3 \Re\left(\frac{k_z^2 - q^2}{\sqrt{k^4 - 4q^2k_\rho^2}}\right).
\end{equation}
Note that in the long-wavelength limit ($q \ll k$), \eqnref{eq:UddaaFT} is the Fourier transform of the instantaneous dipole-dipole interaction. Figure~\ref{fig:UDDFT} shows a plot of $U_{\text{ang}}$. Large spatial frequency ($k\gg q$) corresponds to the instantaneous case, $U_{\text{ang}}(\vector{k}) \simeq -1+3(k_z/k)^2$. At small frequencies there is a singularity at $k^4 = 4q^2 k_\rho^2$, which results in an offset circle in $k$-space, inside of which has a static value of -1.

\section{Numerical Modeling} 
For computational efficiency, we perform simulations in dimensionless units. The dimensionless Hamiltonian from \eqnref{eq:GPE} is
\begin{equation}
  \begin{split}
  \tilde{H} = - \tilde{\nabla}^2 + \left( \tilde{\rho}^2 + \lambda^2 \tilde{z}^2 \right) + 8 \pi \tilde{a} N_{\text{BEC}} \tilde{n}(\vector{r}) \\
  + 8 \pi \frac{\tilde{d}^2}{3} N_{\text{BEC}}\, \mathcal{F}^{-1}_{\text{3D}}\{ U_{\text{ang}}(\vector{k}) \hat{\tilde{n}}(\vector{k})\},
  \end{split}\label{eq:GPEdimensionless}
\end{equation}
where the tilde represents dimensionless quantities with lengths scaled by $l_\rho = \sqrt{\hbar/m \omega_\rho}$, energies by $\hbar \omega_\rho$, and densities by $N_{\text{BEC}}(m \omega_\rho/\hbar )^{3/2}$. Here we have substituted the Fourier-transformed form of the dipole-dipole potential.

The dipole-dipole system's ground state is computed by means of imaginary time propagation (ITP), with a change of variables $t \to \imath \tau$ \cite{ITPAveraging}. The dimensionless ground state BEC order parameter ($\tilde{\psi}$) is found by iterating through imaginary time with
\begin{equation}
  \tilde{\psi}_{i+1} = \tilde{\psi}_{i} - \Delta\tau\, \tilde{H} \tilde{\psi}_{i},
  \label{eq:ITP}
\end{equation}
until $\tilde{\psi}$ converges. Here $\Delta\tau$ is the imaginary time step between $i$ and $i+1$ iterations, and the order parameter is renormalized after every iteration with $\int d\vector{r}\, |\tilde{\psi}|^2 = 1$. A random value is assigned to $\Delta \tau$ each iteration, which eliminates oscillations in the order parameter. We scale $\Delta \tau$ so that a single random step through imaginary time can change the order parameter by at most 5\% of the previous value.

The order parameter and dimensionless Hamiltonian are sampled over a 2D grid ($\rho,z$) with $N_{\text{radial}}$ and $N_{\text{axial}}$ bins in each direction. The first three terms in \eqnref{eq:GPEdimensionless} are the kinetic energy, trapping potential with trap ratio $\lambda = \omega_z / \omega_\rho$, and s-wave scattering energy with scattering length $\tilde{a} = a/l_\rho$. The last term corresponds to the non-local dipole-dipole interactions, which require a Fourier transform. Three-dimensional fast Fourier transforms are often used to calculate the convolution in the dipole-dipole interaction term, but due to the enforced cylindrical symmetry in \eqnref{eq:UddaaFT}, the discrete Hankel Fourier transform (DHFT) is faster \cite{DHFTLemoine,DHFTJohnson, BohnDHFT}. A DHFT calculates the 3D Fourier transform by performing a Fourier transform in the axial direction and a circularly symmetric Hankel transform in the radial direction. Using a DHFT requires sampling the radial direction at Bessel zeros, $j_0(n)$, such that the $i$th radial coordinate is $\rho_i = j_0(i+1)/R$, where $R$ is the maximum radial range, for $i = {0,1,\dots,N_\rho-1}$. The axial sampling is linearly spaced such that the $j$th axial coordinate is $z_j = j \frac{Z}{N_{z}-1}$, for $j = {0,1,\dots,N_{\text{axial}}-1}$. Each iteration, a DHFT transform is performed on $n(\vector{r})$, and an inverse DHFT is performed on $U_{\text{ang}}(\vector{k})\hat{\tilde{n}}(\vector{k})$. While the DHFT calculation is computationally feasible, it is significantly slower than any other step in this calculation.

The computation is dramatically slowed by the singularity in $U_{\text{ang}}$ on the surface $k^4 - 4q^2k_\rho^2 = 0$. Near this surface it is difficult to sample $U_{\text{ang}}$ accurately, but supersampling and averaging near the surface of the singularity does reduce errors significantly \cite{ITPAveraging}. Every bin that corresponds to $\sqrt{(k_\rho -q)^2 + k_z^2} > q$ and $k^4 - 4q^2k_\rho^2 < 1$ is sampled $10^5$ times on a finer grid size and then averaged.

The simulation process starts with a randomized order parameter over the radial range $[j_0(1)R/j_0(N_\rho), R]$ and axial range $[0, Z]$. While computing the Hamiltonian from \eqnref{eq:GPEdimensionless} the order parameter is advanced by \eqnref{eq:ITP} using the previous order parameter, and the process repeats at least 1000 times, until the order parameter converges to its ground state. Typical values for $N_{\text{radial}}$ and $N_{\text{axial}}$ are 163 and 1944 bins respectively. These values are chosen to sample properly the scale set by the laser wavelength, and the axial number of bins is further selected for the quickest DHFT.

Consistency checks are done to ensure the ground-state BEC has a constant chemical potential $\mu \psi = H \psi$. The ground-state order parameter in the absence of dipole-dipole is also compared to the well known theoretical value predicted for trapped BECs. Typically we achieve errors no more than 0.5\% of the density in each bin. Results of these simulations are shown in the next Section.

As ITP is a computationally intensive technique with the dipole-dipole potential, we investigated whether a simpler calculation could reach similar results, namely the variational principle (VP). In VP, the ground-state energy configuration is found by minimizing the energy functional,
\begin{equation}
   E = E_{\text{kin}} + E_{\text{trap}} + E_{\text{s-wave}} + E_{\text{DD}},
\end{equation}
in cloud size $\vector{s}$, using a Gaussian ansatz in the frequency domain, $n(\vector{k}) = (2 \pi)^{-3/2} N_{\text{BEC}} \, \exp(-s_\rho^2 k_\rho^2/4 - s_z^2 k_z^2/4)$. In SI units, the terms in the energy functional are
\begin{align}
   E_{\text{kin}} &= \frac{\hbar^2 N}{2 m} \left[ \frac{1}{s_\rho^2} + \frac{1}{2 s_z^2} \right],\\
   E_{\text{trap}} &= \frac{m N}{2} \left[ \omega_\rho^2 s_\rho^2 + \frac{\omega_z^2 s_z^2}{2} \right],\\
   E_{\text{s-wave}} &= \frac{a \hbar^2 N^2}{\sqrt{2\pi} m s_z s_\rho^2},
\end{align}
and
\begin{equation}
   \begin{split}
   E_{\text{DD}} = \frac{N^2 d^2}{2\pi} \left[ \frac{ - \sqrt{2\pi}}{3 s_z s_\rho^2} + \int_{-\infty}^{\infty} dk_z \int_0^\infty dk_\rho\ k_\rho \right. \\
   \left. \Re \left\{ \frac{k_z^2 - q^2}{\sqrt{k^4 - 4q^2k_\rho^2}}\right\} \exp \left( \frac{-s_\rho k_\rho^2}{2} - \frac{ s_z^2 k_z^2}{2} \right) \right].
   \end{split}
\end{equation}
The integral in $E_{\text{DD}}$ is numerically challenging to perform due to the singularity at $k^4 - 4q^2k_\rho^2 = 0$, and the same technique used for ITP did not provide consistent integration results near the singularity. Though the variational approach does not provide accurate retardation calculations, ITP and VP calculations produce the same results for long wavelengths, where the integral vanishes and leaves only the instantaneous term. For short wavelengths, ITP is the preferred method for calculating the ground state of a BEC with dipole-dipole interactions since it reduces inaccuracies from the singularity by supersampling and averaging.

\section{Results}
\begin{figure}
  \includegraphics[width=\linewidth,clip]{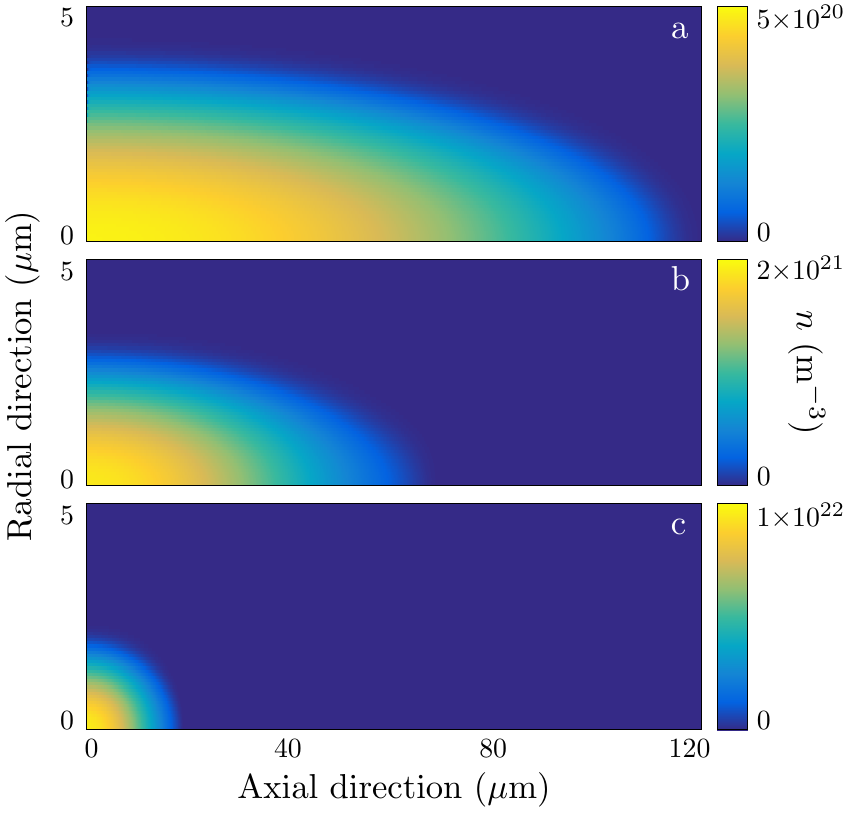}
  \caption{Ground-state BEC density calculated for a 30~$\mu$m inducing laser at different laser intensities: a) $\mathcal{I}=0$, b) $\mathcal{I}=0.66$, and c) $\mathcal{I}=0.92$. As the intensity increases, the interactions are strengthened and the condensate compresses. The axial and radial widths are found by fitting these distributions to a Thomas-Fermi function.}
  \label{fig:Density}
\end{figure}

We calculate the cylindrically symmetric ground-state BEC density using the ITP method described above. We choose parameters to match our existing apparatus, namely a BEC of $1.5\times 10^6$ $^{87}$Rb atoms in a 30:1 pencil-shaped trap, with $\omega_r = 2\pi\times 237$~Hz and $a=100a_0$ \cite{OurParameters}. Using the same initial parameters, we perform simulations varying laser intensity, and thus linearly varying dipole-dipole interaction strength. Figure~\ref{fig:Density} shows the ground-state density of a BEC illuminated by a long-wavelength laser with varying intensities. The increasing dipole-dipole interactions cause the BEC to compress strongly in the axial direction.

\begin{figure}
  \includegraphics[width=\linewidth,clip]{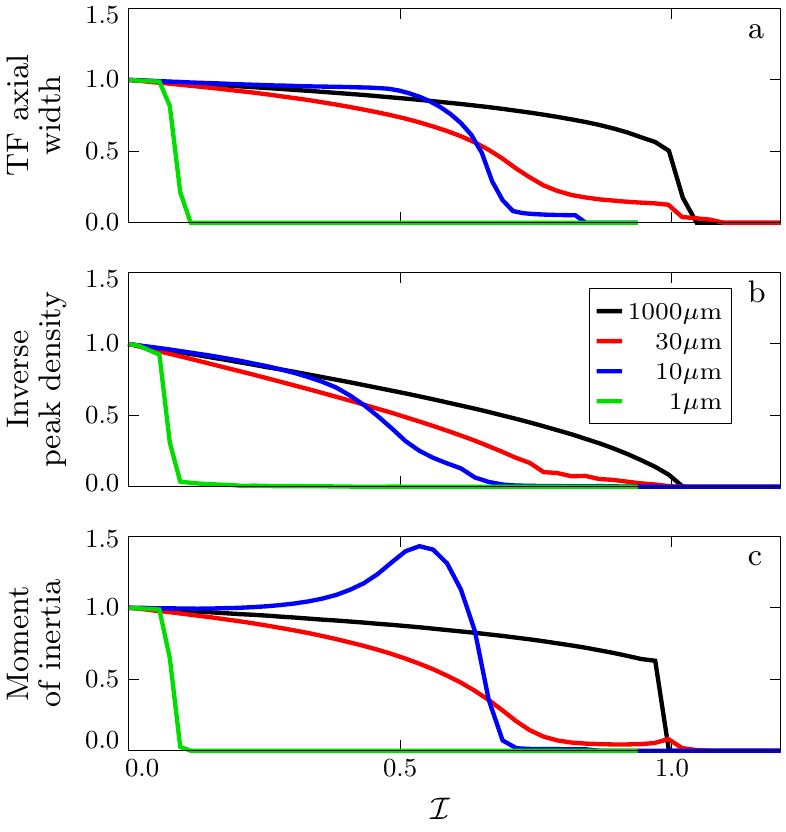}
  \caption{Collapse curves for three relative parameters that indicate collapse -- a) condensate axial size, b) peak density, and c) moment of inertia -- as a function of scaled laser intensity. Each parameter is scaled by their value when dipole-dipole interactions are turned off, $\mathcal{I}=0$. Each plot contains four different laser wavelengths: 1000~$\mu$m (black), 30~$\mu$m (red), 10~$\mu$m (blue), and 1~$\mu$m (green). The critical intensity is chosen to be when the condensate reaches half the unperturbed width or moment of inertia, or ten times the initial density. Despite significant distortions to Thomas-Fermi distributions from compression, these metrics for critical intensity agree well with each other and are deemed suitable for quantifying critical intensity even at short wavelengths.}
  \label{fig:compression}
\end{figure}

The condensate compression as a function of laser intensity is shown in \figref{fig:compression}. The amount of compression is quantified by measuring the peak density, axial width, and moment of inertia and normalizing by the corresponding parameters in the absence of dipole-dipole interactions ($I=0$). The peak density and axial width are determined through fits to a Thomas-Fermi profile, $n(\rho,z) = n_0 [1+ (\rho/s_\rho)^2 + (z/s_z)^2]$, with peak density $n_0$ and axial and radial widths $s_z$ and $s_\rho$. The moment of inertia is calculated using $\sum_{\text{bins}} (\rho^2 + z^2) n(\rho,z) 4 \pi\, \Delta\rho\, \Delta z$. These three different indicators of compression are used to check for consistency, since when the condensate compresses significantly at short laser wavelength, the distribution can depart significantly from a Thomas-Fermi profile.

As the laser intensity increases, the condensate begins to collapse in size. The intensity where this occurs is the critical intensity, $I_{\text{crit}}$. Above this intensity, the condensate shrinks to near zero size from the strongly attractive dipole-dipole interactions, and the mean-field approach breaks down as the interactions become too strong. We choose the critical intensity by determining the intensity at which the axial width or moment of inertia drops to $50\%$ of its initial value, or when the peak density increases tenfold. Despite clear departures from Thomas-Fermi distributions, all three size parameterizations used for consistency checks give the same critical intensity, so henceforth the moment of inertia is used for determining $I_{\text{crit}}$. The critical intensity corresponds to the intensity where the s-wave and dipole-dipole interaction strengths from \eqnref{eq:GPEdimensionless} are equal, $\tilde{a} = \tilde{d^2}/3$ \cite{PermaExact}. At long wavelengths, where no retardation effects are present, the critical intensity is analytic,
\begin{equation}
  I_{\text{LW}} = \frac{12 \pi c \epsilon_0^2 \hbar^2}{m} \frac{a}{\alpha^2},
  \label{eq:Icrit}
\end{equation}
but short wavelengths still require numerical modeling. Within collapse curves, the scaled intensity $\mathcal{I} = I/I_{\text{LW}}$ is used as an atom-independent measure of interaction strength, where full collapse occurs at $\mathcal{I} = 1$ for long wavelengths.

Certain wavelengths also allow for intermediate collapse states, where the axial component significantly compresses to the point that s-wave scattering balances dipole-dipole interactions at an intermediate size. These intermediate states are only found near $\lambda \sim 30\mu m$ for our parameters (see Fig.~\ref{fig:compression}). More generically, these intermediate states can be found at wavelengths that are comparable to the size of the condensate. Only found above critical intensity, these intermediate collapse points are interesting, but do not influence the determination of $I_{\text{crit}}$.

\begin{figure}
  \includegraphics[width=\linewidth,clip]{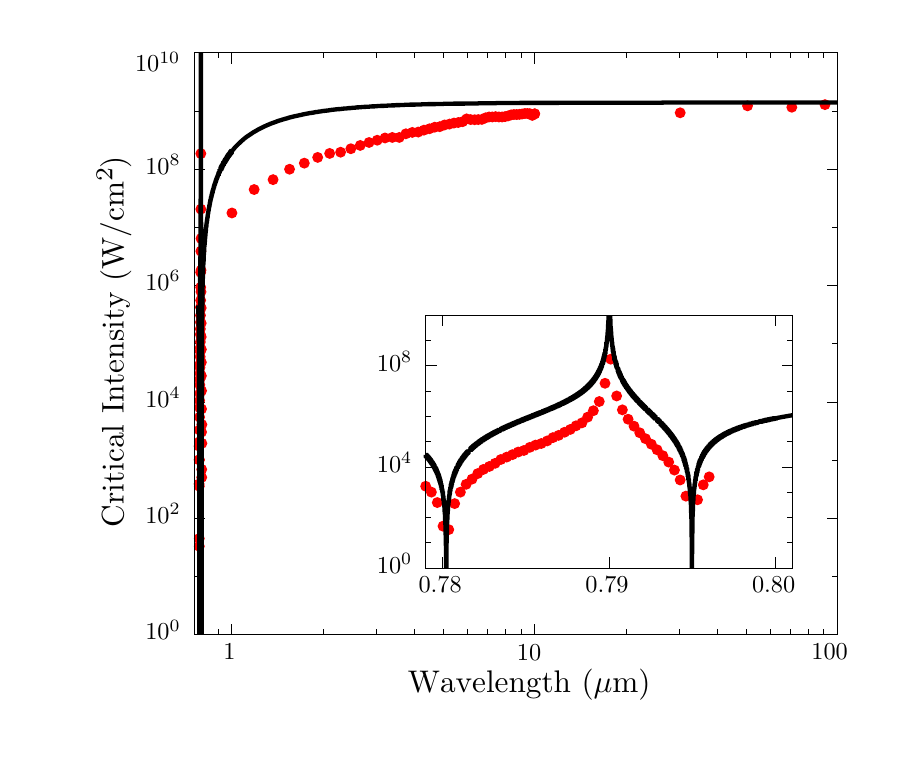}
  \caption{Critical intensity example for $^{87}$Rb with 30:1 trap ratio and $1.5\times10^6$~atoms. The strong dip in critical intensity at lower wavelengths is due to a large increase in polarizability near $^{87}$Rb resonances at 780~nm and 795~nm. The difference between the critical intensity found using the long-wavelength approximation (black) and the ITP-calculated critical intensity (red) is due to retardation effects, and the ratio gives the retardation enhancement. The inset shows the critical intensity near resonance, where critical intensity varies due to the drastic change in polarizability.}
  \label{fig:Icrit}
\end{figure}

To study the wavelength dependence, we perform simulations and obtain collapse curves for a range of wavelengths, and we vary the laser intensity for each wavelength to find the critical intensity. The critical intensity as a function of wavelength is shown in \figref{fig:Icrit}, along with the long-wavelength approximation. Here we see a clear departure from long-wavelength behavior, as the critical intensity from the full ITP calculation is significantly lower at shorter wavelengths. Near-resonance wavelengths benefit from the sharp increase in $\alpha$, as well as retardation effects, leading to critical intensities of $10^4$~W/cm$^2$ and lower. These critical intensities are four orders of magnitude lower than the intensities calculated at long wavelengths and can be supplied with a 1~W laser focused to $\sim 100$~$\mu$m.

\begin{figure}
  \includegraphics[width=\linewidth,clip]{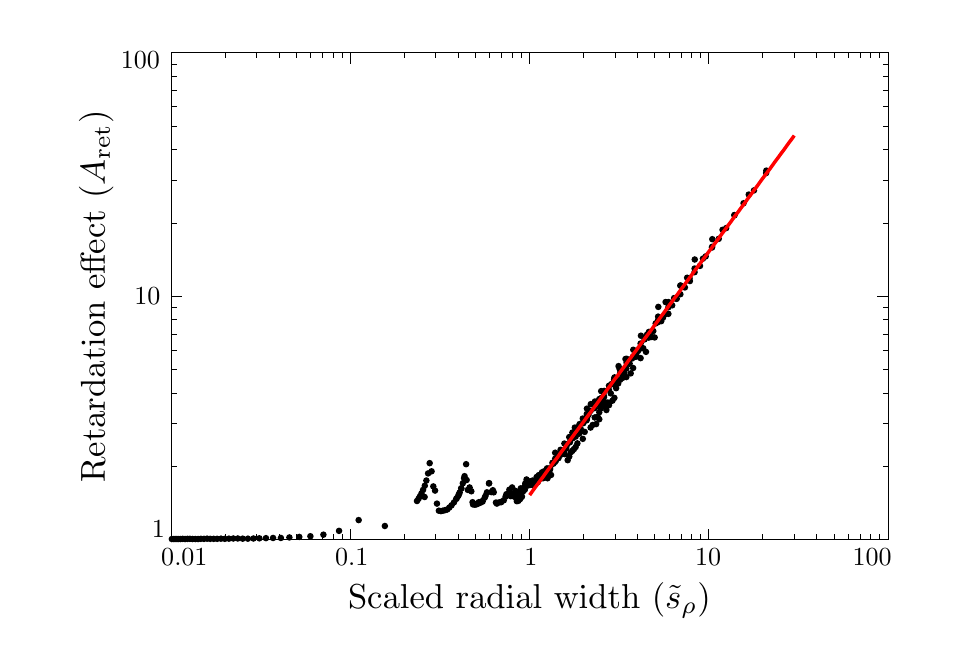}
  \caption{Amplification of dipole-dipole interactions due to retardation effects, $A_{\text{ret}} = I_{\text{LW}}/I_{\text{crit}}$. The dependent parameter is the scaled condensate radial width $\tilde{s}_\rho = s_\rho/\lambda$. A phenomenological linear function is fit to the data above $\tilde{s}_\rho \sim 1$. Retardation effects can amplify dipolar interactions over the instantaneous interaction by as much as 30-fold.}
  \label{fig:RetEffect}
\end{figure}

The deviation from the long-wavelength critical intensity is interpreted as due to retardation effects, quantified by $A_{\text{ret}} = I_{\text{LW}}/I_{\text{crit}}$. The retardation effect is atom-independent, since the size of the effect depends only on the condensate size relative to the laser wavelength. We plot the atom-independent retardation effect in \figref{fig:RetEffect}, fixing axial width and atom number while varying radial width. For long wavelengths ($\lambda \gg s_\rho$) retardation effects are negligible. However, at short wavelengths ($\lambda \lesssim s_\rho$) retardation effects lead to as much as a $30$-fold increase in interaction strength. As seen in Fig.~\ref{fig:RetEffect}, the change in scaling behavior occurs when the laser wavelength becomes smaller than the BEC's radial width, when the BEC is large enough that atomic dipoles on one side of the BEC reside in the extended, non-local region of the potential produced by atoms on the other side.

This increase in interaction strength from retardation effects is due to the dependence on $qr$ in the interatomic interaction [\eqnref{eq:Uddaa}], as well as the long-range nature of the interaction. Although the complicated dipole-dipole potential does not allow for analytic results in the retarded regime, we note that in the large condensate limit ($\tilde{s}_\rho > 1$) retardation effects follow a simple linear relationship over several decades. Thus we phenomenologically fit $A_{\text{ret}}$ with $A_{\text{ret}} \simeq 1.5\frac{s_\rho}{\lambda}$. We can use this model to predict reduced critical intensities for other BECs of atoms with no permanent dipole moment, where the critical intensity is $I_{\text{crit}} \simeq I_{\text{LW}} / A_{\text{ret}}$. For wavelengths that are similar to the condensate radial width ($0.1 < \tilde{s}_\rho < 1$), there are sharp increases in the retardation effect as only a fraction of the condensate's interactions are retarded, and the exact fraction of a wavelength contained within the condensate is important. As the wavelength decreases further ($\tilde{s}_\rho > 1$), the majority of the condensate undergoes retarded interactions leading to the calculated linear tread with dependence on $qr$.

\begin{figure}
  \includegraphics[width=\linewidth,clip]{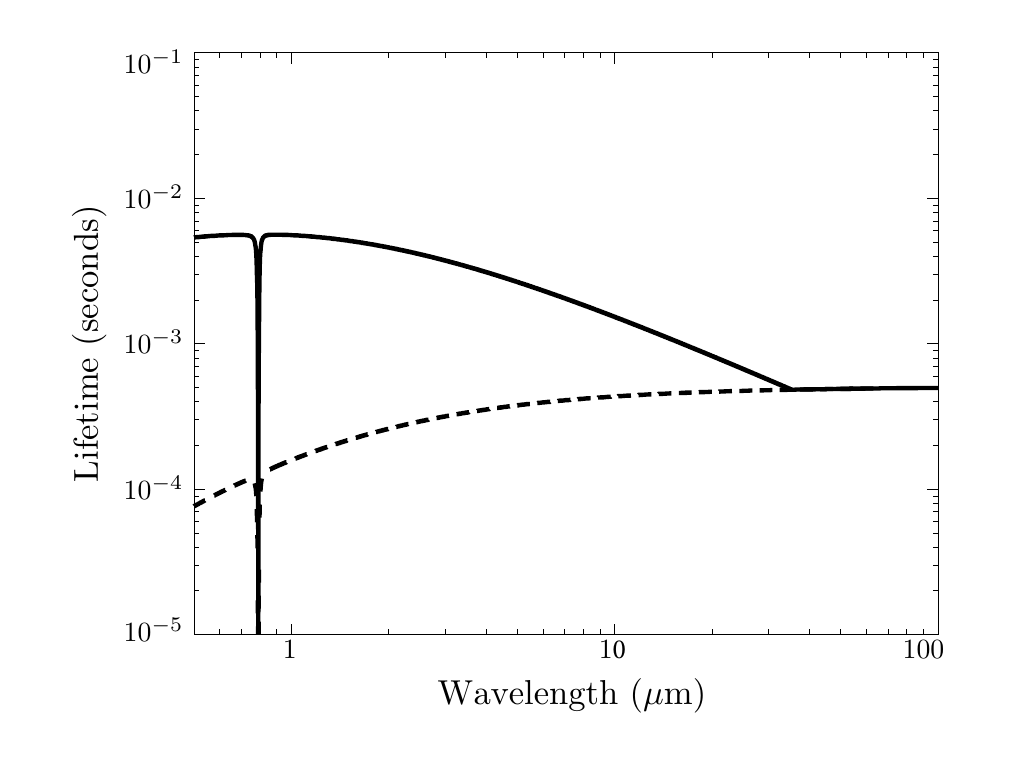}
  \caption{Lifetime of a 10:1 pencil-shaped $^{87}$Rb BEC ($1.5\times10^{6}$ atoms) irradiated by a laser at critical intensity. The lifetime in the long-wavelength limit (dotted) is compared to the lifetime with retardation enhancements included (solid), where the phenomenological fit for the retardation enhancement is used. A 30-fold increase in lifetime is expected near the atomic resonances at $\sim 0.78 \mu$m, due to retardation enhancement.}
  \label{fig:Lifetime}
\end{figure}

Next we consider the feasibility of observing these laser-induced dipole-dipole effects in a BEC. Using the phenomenological equation for retardation effects, coupled with atom-dependent dipole-dipole interaction strengths, we can search for more auspicious wavelengths and atoms. The strongest interactions are found near atomic resonances, where $\alpha$ increases sharply. However, atoms are more likely to absorb photons from near-resonant lasers due to high photon scattering rates, and one would expect a BEC to be destroyed quickly. The scattering-limited lifetime of $^{87}$Rb atoms at the critical intensity is shown in \figref{fig:Lifetime}. Surprisingly, the drop in critical intensity due to increased polarizability and retardation balances the decreased detuning's effect on scattering; thus, atomic lifetimes near resonance actually slightly increase. These short wavelengths have the added benefit of requiring significantly lower laser powers because of the reduced critical intensity.

The best achievable lifetime for a $^{87}$Rb BEC for wavelengths between 750~nm and 10~$\mu$m and at the critical intensity is only $\sim5$~ms. This lifetime is too short to allow a BEC to equilibrate before substantial loss reduces the effect of dipole-dipole interactions, but the lifetime may be increased by reducing the s-wave length scattering using a Feshbach resonance. Reducing s-wave scattering allows higher densities, which enhances the strength of dipole-dipole interactions and lowers the critical intensity [\eqnref{eq:Icrit}].  We expect a lifetime of at least $100$~ms to be necessary to study ground state behaviors, and this demanding requirement necessitates reducing the scattering length $20$-fold via a Feshbach resonance, which is challenging to do over an entire extended sample of $^{87}$Rb. For this reason, we also consider other easily-trapped alkali atoms, such as $^{85}$Rb, $^{133}$Cs, and $^{23}$Na with their typical scattering lengths \cite{Rb85Swave, Cs133Swave, Na23Swave}. We note the longest lifetimes in $^{87}$Rb, with its favorable polarizability, and the other three atomic lifetimes near-resonance are approximately 2~ms. However, $^{87}$Rb has few accessible Feshbach resonances \cite{Rb87Feshbach}, unlike the other three atoms. Using one of these other atoms requires a $50$-fold reduction in scattering length to obtain $\sim100$~ms lifetimes. However, this reduction in scattering length is a more attainable task with $^{85}$Rb, $^{133}$Cs, and $^{23}$Na as these atoms have wider and low-field Feshbach resonances.

\section{Conclusions}
We have studied the effect of retardation in laser induced dipole-dipole interactions in a BEC and shown that retardation effects are strong enough to amplify induced dipole-dipole interactions by at least $30$-fold at short wavelengths. This amplification means an experimental realization requires much lower laser intensities than the long wavelength calculations originally suggested. Successful demonstration of the retardation effects is a first step to creating strong enough long-range induced dipole-dipole forces to create a novel new self-bound BEC with gravitation-like dipole-dipole interactions, and it seems essential to work at short wavelengths.

There are a number of concerns with an experiment to demonstrate retarded dipole-dipole interactions. First, the lifetime of the $^{87}$Rb atoms is very short ($\sim5$~ms) without the use of Feshbach resonances. Using a Feshbach resonance can increase the lifetime to $100$~ms or longer by decreasing the required laser intensity with a challenging $20$-fold reduction in scattering length. Alternative atomic choices are $^{85}$Rb, $^{133}$Cs, and $^{23}$Na, but they require a greater reduction in scattering length due to their slightly lower lifetimes ($\sim 2$ms).  Second, to avoid any spurious size-altering effects, any spatially-dependent dipole forces from gradients in laser intensity must be avoided. This criterion will require an extremely smooth laser intensity profile, meaning larger initial laser intensities. Lastly, lensing effects due to the BEC are not considered, but can cause significant intensity gradients, though typically these would not play a large role in a transversely oriented laser.

Further work could be done to explore shorter wavelengths to search for even larger amplification due to retardation effects. However, this limit requires significant computational resources or an alternate technique. We expect the phenomenological approximation to continue to be valid at readily achievable shorter wavelengths, however. Even stronger enhancement could reduce the critical intensity and lengthen lifetimes further, potentially to a realm where Feshbach resonances are not required to observe the effect of retarded induced dipole-dipole interactions.

The authors thank Dorna Niroomand and Malcolm Kennett for helpful discussions.

\bibliography{DipoleDipolePaper}
\end{document}